\documentclass[journal]{IEEEtran}

\newcommand{\be}{\begin{equation}}
\newcommand{\ee}{\end{equation}}
\newcommand{\bea}{\begin{eqnarray}}
\newcommand{\eea}{\end{eqnarray}}
\newcommand{\df}{{\rm d}}

\ifCLASSINFOpdf
\else
\fi
\usepackage{enumitem}

\usepackage{cite}

\usepackage{url}

\usepackage{graphicx}

\usepackage{amsfonts}

\usepackage{color}

\usepackage{amsthm}

\begin{document}
\title{Information capacity of direct detection optical transmission systems}

\author{Antonio Mecozzi,~\IEEEmembership{Fellow,~OSA, Fellow,~IEEE} and Mark Shtaif,~\IEEEmembership{Fellow,~OSA, Fellow,~IEEE}
%\thanks{Manuscript received \today}% <-this % stops a space
\thanks{A. Mecozzi is with the Department of Physical and Chemical Sciences,
University of L'Aquila, L'Aquila 67100, Italy. M. Shtaif is with the Department of Physical Electronics,
Tel Aviv University, Tel Aviv 69978, Israel. }}

\maketitle

\begin{abstract} We show that the spectral efficiency of a direct detection transmission system is at most 1 bit/s/Hz less than the spectral efficiency of a system employing coherent detection with the same modulation format. Correspondingly, the capacity per complex degree of freedom in systems using direct detection is lower by at most 1 bit.

%These findings show that the Kramers Kronig transmission scheme recently proposed is the optimal one for a given transmission bandwidth when the receiver is allowed to measure only the intensity of the transmitted field.
\end{abstract}


% \fi
\IEEEpeerreviewmaketitle

\section{Introduction} 
\IEEEPARstart{R}{ecently}, the field of optical communications is witnessing a revival of interest in direct detection receivers, which are often viewed as a promising low-cost alternative to their expensive coherent counterparts \cite{Randel_invited,DMT,MarkOFDM,Lowery,Lowery1,Shie,Randel,Petermann,KKOptica,KKOFC,KKOFCpdp,KKJLT}. This process stimulates an interesting fundamental question to whose answer the present paper is dedicated: What is the difference between the information capacity of a direct detection system and that of a system using coherent detection? 

\begin{figure}
	\centering\includegraphics[width=0.95\columnwidth]{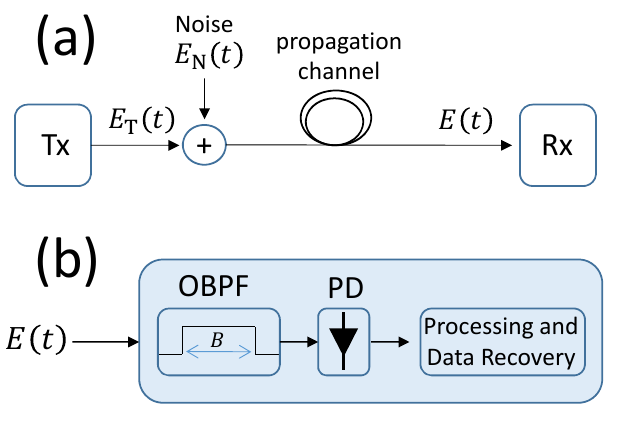}
	\caption{(a) The setup considered in this paper. It consists of an I/Q transmitter (Tx) that can generate any complex waveform whose spectrum is contained in a bandwidth $B$, a stationary noise source of arbitrary spectrum and distribution, a propagation channel, and a receiver (Rx). (b) The schematic of a direct detection receiver. The incoming optical field is filtered with an optical band-pass filter (OBPF) to reject out of band noise and square law detected (without any manipulation of the field) by a single photodiode (PD). The photodiode and subsequent electronics are assumed to have bandwidth of at least $2B$ so as to accommodate the bandwidth of the intensity waveform. }\label{DDRsetup}
\end{figure}

In order to answer this question we consider the channel schematic illustrated in Fig. \ref{DDRsetup}a, which consists of a transmitter that is capable of generating any desirable complex waveform whose spectrum  is contained within a bandwidth $B$, a noise source of arbitrary spectrum and statistics, a propagation channel,
and a receiver. Although the linearity of the channel and the additivity of the noise are immaterial to our analysis, we will assume these properties in the beginning, while postponing the generalization of our discussion to Sec. \ref{Extension}. 
The direct detection receiver in our definition is one that recovers the communicated data from the intensity (i.e. absolute square value) of the received electric field, while using a single photo-diode. As illustrated in Fig. \ref{DDRsetup}b, it consists of a square optical band-pass filter of width $B$ that rejects out of band noise, a photo-diode whose output current is proportional to the received optical intensity\footnote{For simplicity, in what follows we will assume that the proportionality coefficient is 1.}, and a processing unit  that recovers the information. 
%Since in all imaginable scenarios the processing unit operates digitally, it must contain an analog to digital converter (ADC) operating at a minimal sampling rate of  $2B$ in order to avoid any loss of information (as $2B$ is the two sided support of the intensity spectrum). 
The benchmark to which we compare the direct detection receiver, is the coherent receiver, in whose case the complex-valued received optical field is reconstructed.%\footnote{We note that even in the case of a coherent receiver, the complex phase is determined up to an arbitrary constant. See discussion in Sec. \ref{waveforms}.A.} 
 
%An optimal implementation of the coherent receiver is based on the balanced homodyne scheme and it produces two analog outputs containing the real and imaginary parts of the complex envelope of the received field []. Therefore,  digital processing in the homodyne receiver's case requires two ADC units, each operating at a sampling rate of $B$ (the two-sided support of the Fourier transforms of the real and imaginary parts of the complex field). Hence the overall number of samples per second is identical in the cases of coherent and direct-detection receivers.\footnote{Notice that if the coherent receiver were to be implemented by means of a balanced heterodyne scheme (which is also optimal) [], there would be a single analog output that would have to be sampled at the rate of $2B$, once again preserving the number of samples per second.}

Intuitively, it is tempting to conclude that since a direct detection receiver ignores one of the two degrees of freedom that are necessary for uniquely characterizing the electric field, its capacity should be close to half of the capacity of a coherent system. Surprisingly, this notion turns out to be incorrect, and as we show in this paper, the capacity per complex degree of freedom in systems using direct detection is lower by not more than 1 bit than that of fully coherent systems. Correspondingly, the loss in terms of spectral efficiency is  limited to be no greater than 1 bit/s/Hz.

Throughout the paper, in order to simplify the notation, we will assume that the transmitted field is scalar. This assumption does not affect the generality of the results, as the transmission of orthogonal polarization components through linear channels is independent.

%{\color{red} 
%Prior to delving into the formal derivation of this result it is important to stress that contrary to the so called intensity channel considered in \cite{I1,I2,I3}, the transmitter in our case is not constrained in any way and it 

%The problem we discuss in this paper is of interest in the context of the fiber optic channel, where the transmitter usually employs a laser of high coherence, the signal is encoded in the entire complex optical envelope of the field (amplitude and phase) and the optical channel has a limited optical bandwidth. This case should not be confused with the so called intensity channel \cite{I1,I2,I3} where the signal is encoded in the \textit{intensity only of the complex optical envelope}. In such systems, usually, transmitter and the receiver have the same bandwidth. This is a good model of situations where only the intensity of the optical field is modulated like in legacy intensity modulation and direct detection (IMDD) fiber optics systems \cite{PTL} or in free space optical systems where the transmitter is an intensity modulated low-coherence (hence with a large phase noise) light emitting diode.}

%

\section{Relation to prior work}
Our result, stating that a direct detection channel is characterized by almost the same capacity as a coherent channel, requires some clarification in view of its being in an apparent contradiction to prior work, where the capacity of a seemingly similar channel was found to be lower by approximately a factor of two. This work consists of Ref. \cite{PTL} published by the authors of the current manuscript, as well as a number of more recent works, the most relevant of which are contained in Refs. \cite{La,I1,I2,I3}.
In order to avoid confusion, we will adopt the terminology of \cite{I2} and refer to the channels studied in those papers as various flavors of the \textit{intensity channel,} whereas the term \textit{direct detection channel} will only be used in reference to the channel that we study here. The reason for this apparent contradiction boils down to the fact that all the versions of the intensity channel assume that the information is encoded at a given rate $B$ directly onto the intensity of the transmitted optical signal and then it is recovered by sampling the received signal's intensity at exactly the same rate. In all cases, the channel is assumed to be memoryless and the optical bandwidth (and hence also the spectral efficiency) do not play any role. The studies in \cite{La,I1,I2,I3} can be given practical justification when considering very short-reach, or very old optical systems that used a low-coherence optical source (such as a light-emitting diode). Indeed, with such sources the optical phase is far too noisy to be used for transmitting information and the source linewidth is so much greater than the modulation bandwidth that relating to spectral efficiency in the modern sense is not meaningful.   

In contrast to the above, the direct detection channel is inspired by modern fiber-optic communications systems, the vast majority of which relies on a highly coherent laser source -- one whose linewidth is substantially smaller than the bandwidth of the modulation. This is the reason for our assumption that the transmitter in the direct detection channel can encode information into \emph{any} complex waveform with the only constraint being that its spectrum is contained in some bandwidth $B$. In addition, since the process of photo-detection involves frequency doubling, the spectrum of the measured intensity is contained in a two-sided bandwidth of $2B$, and hence sampling at the rate of $2B$ is imperative in order to extract the information present in the photo-detection current. 

In order to further clarify the difference between the direct detection channel and the intensity channel, we denote the complex-valued field received after optical filtering, by $E(t)$. Since the spectrum of $E(t)$ is contained in a bandwidth $B$, it can be rigorously expressed as  
\be E(t) = \sum_{k=-\infty}^{\infty} E_k  \mathrm{sinc} \left[\pi (Bt-k) \right], \label{NS} \ee
where $\mathrm{sinc}(x)=\sin(x)/x$, and where $E_k=E(t=k/B)$ are the complex-field samples carrying  the transmitted information. The detected photo-current is proportional to $I=|E(t)|^2$. If this photocurrent were to be sampled at a rate of $B$, as in \cite{PTL,La,I1,I2,I3}, the samples at $t=n/B$ would have been equal to $I_n=|E_n|^2$, and the phase information would have been lost. In this case the drop in the amount of extracted information (and hence the capacity) would have been roughly a factor of 2, similarly to the results obtained in \cite{PTL,La,I1,I2,I3}. Yet, in the direct detection channel the sampling of the photo-current is done at a rate of $2B$, so that the middle-point samples that are taken at $t=(n+0.5)/B$ are also obtained. These middle samples are given by
\bea I_{n+1/2} &=& \sum_{k,m=-\infty}^{\infty} \mathrm{sinc} \left[\pi \left(n-m +\frac 1 2\right) \right] \nonumber \\
&& \times \mathrm{sinc} \left[\pi \left(n-k+\frac 1 2\right)\right] E_{m}^*E_{k}, \eea
and they are clearly affected by the phase differences between the various complex-field samples. In fact, as our final result indicates, knowledge of all intensity samples ($I_n$ and $I_{n+1/2}$ for all $n$) allows one to collect almost all of the information contained in the complex optical field.

We note that the idea of increasing the information rate by sampling the received analog signal at a rate that is higher than $B$, has been considered previously \cite{Gilbert}. This is a natural idea in cases where the receiver contains a nonlinear element that expands the analog bandwidth of the received waveform so that sampling at $B$ is no longer sufficient in order to collect all the information from the analog waveform. In our case the nonlinearity is that of square-law detection and it expands the analog bandwidth by exactly a factor of 2. Hence, unlike the case studied in \cite{Gilbert}, where the doubling of the sampling rate produced only minuscule benefits, here sampling at $2B$ is sufficient in order to extract all the information contained in the analog intensity waveform, and there is no benefit in increasing the sampling rate farther.

Finally, it is instructive to relate to the most widespread example, where the additive noise of Fig. 1a is white Gaussian. In this case, our theory implies that the capacity of the direct detection channel is within 1 bit of $\log(1+\mathrm{SNR})$ \cite{Shannon}, where the SNR is the ratio between the average power of the information carrying signal and the variance of the filtered noise (summed over both quadratures). Conversely, as demonstrated in \cite{PTL} and \cite{I3}, the capacity of the intensity channel (i.e. one that samples the received intensity at the rate $B$) in the limit of high SNR, is  $0.5\log(\mathrm{SNR}/2)$ --- roughly half of the direct detection channel's capacity.\footnote{In Ref. \cite{PTL}, the capacity in the high SNR limit is written as $0.5\log(\mathrm{SNR}/4)$, but the difference is only in the SNR definition, which relates to the noise variance in one quadrature.}

\section{The information capacity of a direct detection receiver}\label{waveforms}
\subsection{The definition of distinguishable waveforms}
Usually, in engineering practice, two waveforms $E_1(t)$ and $E_2(t)$ are said to be distinguishable when the energy of the difference between them is greater than 0,
\be \int_{-\infty}^\infty |E_1(t)-E_2(t)|^2\df t>0.\label{D1}\ee
In the context of optical communications this definition is  too restrictive, because in all cases of interest, optical receivers (including coherent receivers) are unable to distinguish between waveforms that differ only by a constant (i.e. time independent) phase.\footnote{In order to overcome this limitation, the transmitter and reciever would have to share an exact time-reference (on the scale of a fraction of a single optical cycle!). In principle, this can be achieved by means of an atomic clock. However, the costs of such a solution on the one hand, and the minuscule potential benefit in terms of information rate, on the other hand, ensure that this solution isn't deployed.} Owing to this reality, we define waveforms to be distinguishable only when they can be told apart by an ideal coherent receiver. Formally, this means that $E_1(t)$ and $E_2(t)$ are distinguishable only when they remain distinguishable according to Eq. (\ref{D1}), even if one of them is rotated in the complex plane by some arbitrary constant phase, i.e. 
\be \int_{-\infty}^\infty |E_1(t)-e^{i\theta}E_2(t)|^2\df t>0.\label{different}\ee
for all values of $\theta$. Notice that distinguishability by means of a coherent receiver (\ref{different}) doesn't necessarily imply distinguishability by means of a direct detection receiver. The gap between the two is the subject of the subsection that follows.

%In all subsequent mention of the term `distinguishable' or `different' with respect to waveforms, the definition introduced in Eq. (\ref{different}) will be used.
\subsection{The multiplicity of complex waveforms having the same intensity}
We consider a complex signal $E(t)$, whose spectrum is contained within a bandwidth $B$ and which is periodic in time, with a period $M / B$, with $M$ being an integer. The assumption of periodicity is not a limiting factor in our arguments, as once the main results are established, the non-periodic case can be addressed by assuming the limit of $M\to\infty$. A direct detection receiver can only exploit the intensity $I(t)=|E_0(t)|^2$ in order to extract the transmitted data. Our first claim, which is key to proving the main arguments of this paper, is that there are at most $2^{M-1}$ distinguishable legitimate waveforms $E_j(t)$ (with $j=0,1,\dots,2^{M-1}$), whose intensities $|E_j(t)|^2$ are equal to $I(t)$. 
%We should clarify that our definition of difference between waveforms ignores constant phase. In other words, two waveforms $E_j(t)$ and $E_k(t)$ are said to be distinguishable if and only if $\int|E_j(t)-\exp(i\theta)E_k(t)|^2\df t>0$ for all values of $\theta$. 
An illustration of this idea in the case of $M=4$ can be found in Figure 2.

\begin{figure}
	\centering\includegraphics[width=0.95\columnwidth]{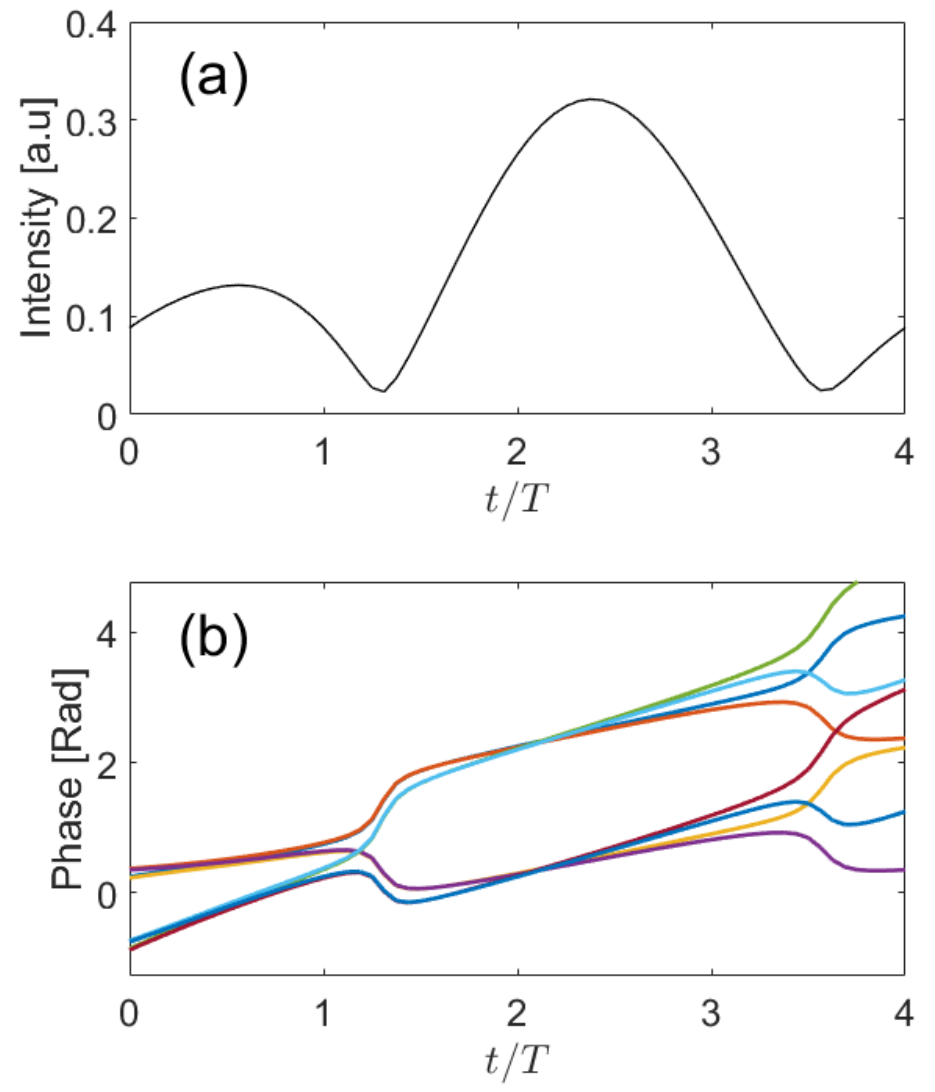}
	\caption{An example of 8 different waveforms in the case of $M=4$ (so that $2^{M-1}=8$), all band-limited to $[0,B]$ (with $B=1/T$) and all having the intensity shown in (a). In (b) the phases $\phi(t)$ of these eight waveforms are plotted. As explained in the text, this is the largest number of distinguishable waveforms that are band-limited to $[0,B]$ and have the same intensity. We stress that waveforms that only differ by a constant (time-independent) phase are not counted as distinguishable in our definition.}\label{Waveforms}
\end{figure}

In order to formally prove our statements, we express $E(t)$ as a Fourier series having at most $M$ non-zero elements,  
\be E(t) = \sum_{k=0}^{M-1} F_k e ^{- i k \Omega t }, \label{FS} \ee 
where $\Omega =  \frac{2 \pi B}{M}$. The Fourier coefficients are given by
\be F_k = \frac{B}{M}\int_0^{M/B}E(t)e^{ik\Omega t}\df t=\frac1 {M} \sum_{n=0}^{M-1} E_n e^{i \frac{2 \pi k n}{M}},\label{Xk}\ee
with $E_n=E(t=n/B)$, and where the second equality in (\ref{Xk}) takes advantage of the relation between the Fourier series coefficients and the discrete Fourier transform in band-limited periodic signals.  We now assign 
\be A(Z) = \sum_{k=0}^{M-1} F_k Z^k \label{ZS} \ee 
to be the Z-transform of the Fourier coefficients $\{F_k\}_{k=0}^{M-1}$. Clearly, $E(t)=A\big(\exp(-i\Omega t)\big)$, and hence special attention needs to be paid to the cases where the value of $Z$ is on the unit circle. When $F_{M-1}\neq 0$,  $A(Z)$ is an $M-1$ degree polynomial which admits $M-1$ zeros, and it can be expressed as
\be A(Z)=C\prod_{k=0}^{M-1}(Z-Z_k),\hspace{0.1cm}\mathrm{with}\hspace{0.2cm} C=\frac{F_0}{\prod_{k=0}^{M-1}(-Z_k)}.\ee

Consider now the functions 
\be U_k(Z)=\frac{ZZ_k^*-1}{Z-Z_k},\label{UkZ}\ee
one for each zero of $A(Z)$. Since $|U_k\big(\exp(i\theta)\big)|=1$, these functions have the property that the action of $U_k\big(\exp(-i\Omega t)\big)$ on $E(t)$ produces a pure phase modulation and hence they can be considered as the dual of \emph{all-pass filters} where time and frequency are interchanged. When any combination of these functions multiplies $A(Z)$, it does not change the degree of the resulting polynomial, as the corresponding zeros of $A(Z)$ are simply reflected with respect to the unit circle, as illustrated below. For example, if we multiply $A(Z)$ by the product of $U_1(Z)$, $U_3(Z)$ and $U_7(Z)$, the zeros $Z_1$, $Z_3$, and $Z_7$ are replaced by $1/Z^*_1$, $1/Z^*_3$, and $1/Z^*_7$, respectively. Yet, the modulus of the product $A(Z)U_1(Z)U_3(Z)U_7(Z)$ remains identical to the modulus of $A(Z)$ when $Z$ is on the unit circle (and in particular, when $Z=\exp(-i\Omega t)$). Since there is a total of $M-1$ functions $U_k(Z)$, there are $2^{M-1}$ functions $A_j(Z)$ that have the same modulus on the unit circle. Thus, we end up with $2^{M-1}$ time waveforms $E_j(t)=A_j\big(\exp(-i\Omega t)\big)$ whose intensity is $I(t)$, i.e. identical to the intensity of $E(t)$. Note that $\{U_k(Z)\}_{k=0}^{M-1}$ are the only functions applying a pure phase modulation to $E(t)$ that also preserve the number of elements in the Z-transform (the degree of the polynomial) and consequently the spectral width of the resulting time waveforms. For this reason  $\{E_j(t)\}_{j=1}^{2^{M-1}}$ are the \emph{only} temporal waveforms whose intensity is $I(t)$, and whose spectrum is fully contained within the bandwidth $B$. 
A further discussion of the uniqueness of the waveforms $\{E_j(t)\}_{j=1}^{2^{M-1}}$ is provided in the appendix. 

Prior to concluding this section, it is interesting to stress that while $2^{M-1}$ is the highest possible number of distinguishable band-limited waveforms whose intensity equals $I(t)$, the actual number of such waveforms is $2^{N_0}$, where $N_0$ 
is the number of zeros that are not located on the unit circle. That is because when a zero $Z_l$ falls on the unit circle, $U_l(Z)$ can be easily verified to be  a constant (i.e. $Z$-independent) phase-factor whose application to $A(Z)$ does not produce a new waveform. Note also that in this situation $E(t_l)=0$ with $t_l=\arg(Z_l)/\Omega$.\footnote{Interestingly, when $E(t)$ equals 0 exactly $M-1$ times within the time period  $M/B$, then it is also the only waveform that is band-limited to $B$ and has that particular intensity.} 

\subsection{The implications to capacity}\label{Sec3}
We now prove the following relation between $C_d$ -- the information capacity of the direct detection channel, and the capacity $C_c$ of a system using coherent detection 
\be C_c-1\leq C_d\leq C_c\label{Cap},\ee
where in all cases, we are referring to the capacity per complex degree of freedom.\footnote{The number of complex degrees of freedom is $M$, which is the product of the temporal duration of the signal $M/B$ and the bandwidth $B$.}
We denote by $X$ the input alphabet of our channel
and by $Y'$ the output alphabet available to a coherent receiver. The output alphabet that is available to a direct detection receiver is denoted by $Y$. Since no constraints are imposed on the transmitter, the alphabet $X$ contains all complex waveforms without restriction. The alphabet $Y'$, on the other hand, contains only those complex waveforms that are band-limited to $B$, whereas $Y$ contains all real-valued waveforms that can be obtained by squaring the absolute value of the waveforms contained in $Y'$. Communication requires that a probability $p_x(x)$ is prescribed to the transmission of each individual waveform $x\in X$.\footnote{Since $x$ is an element of the alphabet $X$ it represents a time dependent waveform. Nonetheless, in order to keep the notation simple, we avoid writing $x(t)$, leaving the time dependence of $x$ implicit. Additionally, in order to avoid over-cluttering the notation, we denote the probability distribution of $x$ simply by $p_x(x)$. A similar practice is used with the elements of $Y$ and $Y'$.} The effect of the communications channel (noise distortions etc.) is characterized by the conditional probabilities of detecting a given element $y'\in Y'$ (in the case of coherent detection), or $y\in Y$ (in the case of direct detection), given that a particular element $x\in X$ was transmitted. These conditional probabilities are denoted by $p_{y'|x}(y'|x)$ and $p_{y|x}(y|x)$, respectively. The mutual information per complex degree of freedom between the transmitter and each of the two receivers equals \cite{Shannon}
\bea I(X;Y') &=& \frac 1 M \left[H(Y')-H(Y'|X)\right],\label{I11}\\
I(X;Y) &=& \frac 1 M \left[H(Y)-H(Y|X)\right], \label{I2} \eea
where the entropy $H(Y)$ and the conditional entropy $H(Y|X)$ are given by\footnote{In line with our simplified notation, summation over $x$ and $y$ should be interpreted in a generalized sense. In addition $p_y(y)=\sum_x p_x(x)p_{y|x}(y|x)$.} 
\bea \hspace{-0.1cm}H(Y)\hspace{-0.3cm} &=&\hspace{-0.3cm} -\sum_y p_y(y)\log_2\Big(p_y(y)\Big) \label{entr1}\\
\hspace{-0.1cm}H(Y|X)\hspace{-0.3cm}&=&\hspace{-0.3cm}-\hspace{-0.1cm}\sum_x\hspace{-0.05cm} p_x(x)\sum_y\hspace{-0.05cm} p_{y|x}(y|x)\log_2\Big(p_{y|x}(y|x)\Big),\label{entr2}\eea
and where the corresponding equations $H(Y')$ and $H(Y'|X)$ are obtained by replacing $y$ with $y'$ in all places. The capacities $C_c$ and $C_d$ are obtained by maximizing the mutual information values of Eqs. (\ref{I11}) and (\ref{I2}) with respect to the transmitted distribution $p_x(x)$.
{In order to derive Eq. (\ref{Cap}), we take advantage of the relation 
\bea I(X;Y') &=& I(X ; Y', Y) = I(X ; Y) + I(X ; Y' | Y) \nonumber \\
&\le& I(X ; Y) +  \frac {M-1} M, \label{I1}\label{Is} \eea
where the first equality follows from the fact that $I(X ; Y | Y') = 0$, and the second equality follows from the relations 
%
%\bea I(X ; Y) &=& \frac1 M \left[H(X;Y',Y) - H(X,Y'|Y)\right], \\
%I(X ; Y' | Y) &=& \frac1 M \left[H(X,Y'|Y) - H(X|Y',Y)\right], \eea
%
%
\bea I(X ; Y', Y) &=& \frac1 M \left[H(X) - H(X|Y',Y)\right], \\
I(X ; Y) &=& \frac1 M \left[H(X) - H(X|Y)\right], \\
I(X ; Y' | Y) &=& \frac1 M \left[H(X|Y) - H(X|Y',Y)\right]. \eea
The last inequality is true because $Y' $ can take no more than $2^{M-1}$ functional values for any given $Y$. %\footnote{{\color{red} The second equality holds because $I(X ; Y', Y) = [H(X;Y',Y) - H(X|Y',Y)]/M$ is equal to the difference between $I(X ; Y) = [H(X;Y',Y) - H(X,Y'|Y)]/M$ and $I(X ; Y' | Y) = [H(X,Y'|Y) - H(X|Y',Y)]/M$.}}

In the limit of large $M$, Eq. (\ref{Is}) reduces to}
\be I(X;Y')-1\leq I(X;Y)\leq I(X;Y'). \label{Is1}\ee
Note that expressions (\ref{Is}) and (\ref{Is1}) hold for any distribution of the transmitted alphabet $P_X(x)$. This means that, for any modulation format, the information per complex degree of freedom that can be extracted when using a direct detection receiver is at most one bit less than the information per channel use that can be extracted with coherent detection. 

When $p_x(x)$ is set to be the distribution that maximizes $I(X;Y')$, we arrive at 
\be C_c-1\leq I_{p}(X;Y)\leq C_c,\label{I1}\label{C2}\ee
where $I_{p}(X;Y)$ is the mutual information $I(X;Y)$ that corresponds to the distribution $p_x(x)$ for which $C_c$ is attained. Clearly, $C_d\geq I_p(X;Y)$ and hence $C_d\geq C_c-1$. Nonetheless, $C_d$ remains smaller or equal to $C_c$, as follows from the right-side inequality of Eq. (\ref{Is}). This concludes the proof of Eq. (\ref{Cap}). 

Finally, we note that the capacity per complex degree of freedom, which we have evaluated in Sec. \ref{Sec3} is identical to the spectral efficiency, which is the more commonly used term in the context of fiber communications. Hence the spectral efficiency of a direct detection system is at most 1 bit/sec/Hz smaller than that of a system using coherent detection. In order to see that the two are exactly the same, note that $B$ is both the bandwidth of the optical signal as well as the number of complex degrees of freedom that are transmitted per second.

\section{Extension to Nonlinear systems or to non-additive noise}\label{Extension}
Fiber-optic communications are often affected by the nonlinear propagation phenomena taking place in optical fibers. Their effect is not only to distort the signal itself, but also to cause a nonlinear interaction between the signal and noise, in which case the noise can no longer be modeled as additive. From the standpoint of our current study, the only difficulty that is imposed by this situation is that it is impossible to relate to the spectrum occupied by the data-carrying signal as a constant, and hence the definition of spectral efficiency becomes problematic. Nonetheless, it must be stressed that our analysis of the received waveforms  in Sec. \ref{waveforms} did not explicitly assume anything about the type of noise, or propagation. Therefore, our results with respect to the capacity of the optically filtered signal in Fig. 1b, remain perfectly valid. In other words, after band-pass filtering, the information per degree of freedom that is contained in the received complex optical signal is at most one bit larger than the information contained in its intensity. With this said, it must be noted that we do not claim that positioning of a square filter in front of the receiver is an optimal practice in the nonlinear case. Nonetheless, in practical situations encountered in fiber communications, the inclusion of such a filter is practically unavoidable. 

\section{Discussion}

While Eq. (\ref{Cap}) corresponds to the only relevant case of $M\to\infty$, the opposite limit of $M=1$, which can be deduced from Eq. (\ref{Is}) may challenge one's physical intuition, as it predicts equality between the mutual information values corresponding to direct and coherent detection. For this reason the discussion of this special case is interesting in spite of the fact that it is of no practical importance whatsoever. In order to resolve this apparent conundrum, note that the case  $M=1$ represents a situation in which the complex field $E$ is time independent. In particular, the phase difference between any two possible fields is also time independent, implying that the fields are distinguishable only provided that their intensities differ. Hence, in this artificial situation, the coherent receiver has no advantage over the direct detection receiver and therefore their capacities are identical.

Another curious point related to the assumption of periodicity is that it is not the only convenient choice for arriving at the result of  Sec. \ref{waveforms}.B.  
Since $E(t)$ is band limited and its spectrum is contained within $\omega\in[0,B]$, it can be written as 
\be E(t) = \sum_{n=-\infty}^{\infty} E_n e^{- i (B t-n) \pi } \mathrm{sinc} \left[\pi (Bt-n) \right], \label{NS} \ee
where, as noted earlier, $E_n=E(t=n/B)$, and where $\mathrm{sinc}(x)=\sin(x)/x$. If we impose the requirement that $E_n=0$ for $n<0$ and for $n\geq M$, we end up with a band-limited, but non-periodic $E(t)$. Nonetheless, the number of waveforms $E_j(t)$ whose intensity equals that of $E(t)$ remains at most $2^{M-1}$. In order to see that, consider a time interval of $M'/B$ that contains the interval $M/B$ at its center. Assume also that $M'\gg M$, so that the tails of the various sinc functions decay to the extent that the signal within $M'/B$ can be extended periodically without introducing any bandwidth broadening. We may now apply the reasoning of Sec. \ref{waveforms} to the signal in the interval $M'/B$, according to which the number of equal intensity waveforms is 2 to the power of the number of zeros in $A(Z)$ that do not coincide with the unit circle. Evidently, the number of such zeros is at most $M-1$, because there are at least $M'-M$ zeros that fall on the unit circle. These are the zeros of $E(t)$ at the times $t_l=l/B$ (with $l$ being outside of the range of $0$ to $M-1$), which correspond to zeros in $A(Z)$ at $Z_l=\exp(i\Omega t_l)$, i.e. on the unit circle. 

Finally, it is important to stress the consequences of our definition of direct detection, which requires that the incoming optical signal is detected by a single photo-diode per polarization, and without any manipulation of the signal prior to photo-detection.\footnote{The requirement that there is no manipulation of the signal prior to photo-detection can be replaced by the requirement that no manipulation other than all-pass filtering (e.g. dispersion) is applied prior to photo-detection. The reason is that all pass filtering can also be done at the transmitter and hence it does not affect the assumption of this work.} This definition excludes not only the use of a local oscillator, as in coherent detection, but also all self-coherent schemes, such as the ones proposed in \cite{Liu}, and phase-reconstruction schemes of the kind considered in \cite{Shechtman,Yonina}.

\section{Acknowledgement} A. Mecozzi acknowledges financial support from the Italian Government under project INCIPICT. M. Shtaif acknowledges financial support from Israel Science Foundation (grant 1401/16).

\begin{appendices}

\section*{Appendix A} \label{appB}
Our discussion in Sec. \ref{waveforms} involved the statement that the number of distinguishable complex waveforms that are characterized by a bandwidth  $B$ and a period $M/B$, cannot be greater than $2^{M-1}$.
%The justification of this claim was included in the main text. After all, there can be only $M-1$ all-pass filters complying with Eq. (\ref{UkZ}), and the application of any all-pass filter that does not comply with it, would inevitably increase the number of poles and zeros, and hence also the signal's spectral support. 
One justification for this claim is that the only functions in Z-domain that produce pure phase modulation and do not increase the order of the polynomial $A(Z)$ (and hence they do not increase the bandwidth of $E(t)$) are of the form $c_k (Z \tilde Z_k - 1)/(Z_k-1)$, where $Z_k$ is one of the zeros of $A(Z)$ and $c_k$ a constant.
In order for these functions not to change the time-domain intensity waveform of $E(t)$, one must have  $\tilde Z_k = Z_k^*$ and $|c_k|=1$, so that the amplitude of the function's transfer function on the unit circle is 1. This means that the only functions that can be applied to $A(Z)$ without changing neither the order of the polynomial, nor the intensity waveform, are the $M-1$ functions specified in Eq. (\ref{UkZ}). Indeed, the number of such function combinations does not exceed $2^{M-1}$. Here, we also present an alternative proof that is based on the uniqueness of minimum-phase functions.

%The justification of this claim was included in the main text, and it is based on the consideration that the order of the polynomial $A(Z)$ is preserved only if it is multiplied by products functions like $c_k (Z \tilde Z_k - 1)/(Z_k-1)$, where $Z_k$ is one of the zeros of $A(Z)$ and $c_k$ a constant, and that the only way that these functions become an all pass filter when $Z$ is restricted to the unit circle is that $\tilde Z_k = Z_k^*$ and $|c_k|=1$. Here, we also present an alternative proof that is based on the uniqueness of minimum phase functions. 

Given a periodic band-limited waveform, we have shown that by reflecting any of its zeros with respect to the unit circle, as described in Sec. \ref{waveforms}.B, one obtains a different waveform having the same intensity, bandwidth, and time-period. %As the number of zeros is $M-1$, the maximum number of waveforms that can be produced by means of this procedure is $2^{M-1}$. 
If we look at an arbitrary given waveform  $E_0(t)=\sqrt{I(t)}\exp(i\phi_0(t))$ of the above specified characteristics, and identify all of its zeros, we can chose to reflect only the zeros that are inside the unit circle, thereby producing a new waveform  $E_m(t)=\sqrt{I(t)}\exp(\phi_m(t))$, having the same intensity $I(t)$, but a different phase. 
Since the spectrum of $E_m(t)$ is contained between $0$ and $B$, and since all of its zeros in Z-domain are outside the unit circle, it belongs to a special class of functions that is famously known as functions of \textit{minimum-phase} \cite{MinimumPhase}. One of the most well known properties of such functions is that up to an immaterial constant, their phase is uniquely determined by their intensity by means of the the Hilbert transform. Namely,  $\phi(t)=\mathbf{H}\Big[\log \sqrt{I(t)}\Big]+c$, where $\mathbf H[\cdot]$ designates the Hilbert transform and where $c$ is an unknown constant. Since waveforms differing only by a constant phase are indistinguishable in our definition (see Sec. \ref{waveforms}.B), we conclude that the minimum phase function that corresponds to a given intensity profile is unique. 

The uniqueness of the minimum-phase function implies that each waveform in the set of distinguishable equal intensity waveforms that are band-limited to $B$ and periodic in $M/B$, can be obtained from any other waveform in the set by means of functions of the form given in Eq. (\ref{UkZ}), whose effect is to reflect the zeros of the waveforms that it acts upon in Z-domain. Had it not been so, different waveforms in the set would have produced different minimum phase functions. Therefore the total number of distinguishable waveforms in the set cannot exceed $2^{M-1}$.

\end{appendices}

\end{document}